\documentclass[aps,prl,a4paper,twocolumn,superscriptaddress,showkeys,longbibliography]{revtex4-2}

\usepackage[colorlinks=true,bookmarks=false,citecolor=blue,linkcolor=red,hyperfootnotes=true,urlcolor=blue]{hyperref}
\usepackage{dcolumn}
\usepackage{bm}
\usepackage{physics}
\usepackage{graphicx}% Include figure files
\usepackage{dcolumn}% Align table columns on decimal point
\usepackage{bm}% bold math
\usepackage{wasysym}
\usepackage{lipsum}
\usepackage{enumitem}
\usepackage{booktabs} 
\usepackage{gensymb}
\usepackage{xcolor}
\usepackage{amssymb}
\usepackage{xcolor}
\usepackage{amsmath}
\usepackage{mathtools}
\usepackage{soul}
\usepackage{float}

\usepackage{changes}

\definecolor{darkgreen}{RGB}{0,100,0}

\begin{document}

\title{Unidirectional subsystem symmetry in a hole-doped honeycomb-lattice Ising magnet}

\author{Sambuddha Sanyal} \thanks{These authors contributed equally: Sambuddha Sanyal, Alexander Wietek} \email{sambuddha.sanyal@iisertirupati.ac.in}
 \affiliation{Department of Physics, Indian Institute of Science Education and Research (IISER) Tirupati, Tirupati 517507, India}
\author{Alexander Wietek}\thanks{These authors contributed equally:  Alexander Wietek, Sambuddha Sanyal}\email{awietek@pks.mpg.de}
\affiliation{Center for Computational Quantum Physics, Flatiron Institute, 162 Fifth Avenue, New York, NY 10010, USA} 
\affiliation{Max Planck Institute for the Physics of Complex Systems, N\"othnitzer Strasse 38, Dresden 01187, Germany}
\author{John Sous} \thanks{Author to whom correspondence should be addressed} \email{sous@stanford.edu}
\affiliation{Department of Physics, Stanford University, Stanford, CA 93405, USA}

\date{\today}

\begin{abstract}
We study a model of a hole-doped collinear Ising antiferromagnet on the honeycomb lattice as a route toward  realization of subsystem symmetry. We find nearly exact conservation of dipole symmetry verified both numerically with exact diagonalization (ED) on finite clusters and analytically with perturbation theory.  The emergent symmetry forbids the motion of single holes -- or fractons -- but allows hole pairs -- or dipoles -- to move freely along a one-dimensional line, the antiferromagnetic direction, of the system; in the transverse direction both fractons and dipoles are completely localized. This presents a realization of a `unidirectional' subsystem symmetry.  By studying interactions between dipoles, we argue that the subsystem symmetry is likely to continue to persist up to finite (but probably small) hole concentrations.
\end{abstract}

\maketitle

\emph{Introduction}. Fractons are the newest addition in the line of exotic quasiparticles in condensed matter physics. Recent years have witnessed tremendous progress in understanding fractons~\cite{CXU0,chamon_2005_topoglass,bravyi_topo_exact3d,Haah_code,castelnovo_chamon_topoglass,pclock_nussinov,Yoshida_fractalqsl,Vijay_haah_fu_2015,Vijay_haah_fu_2016,Kim_Haah_loc,Williamson_fractal,Pretko_U1_2017,Pretko_EM_2017,Coupled_layers_Hermele,Hseieh_fracton_parton,Balents_coupledspinchain,Kevin_kim_nntwospin,Petrova_3Dqsl,Pretko_finitetemp_u1,Pretko_emergent_gravity,Prem_glassy_translationinvariant,Prem_Pretko_emergent,Devakul_correlation,Emergent_conservation_schmitz,CageNet,HYan,Sous2020Fracton1,Sous2020Fracton2,SF_emergent_dipole_conservation_Yuan,DefectnetworkTopo,Hspace_shattering_khemani,Coupledwire_sullivan,SlagleFFT,CXU1,CXU2,CXU3,Jensen1,Jensen2,fracton_review_hermele,PretkoReview}, which unmasked  connections with other areas of physics including  topological order~\cite{chamon_2005_topoglass,Haah_code,Vijay_haah_fu_2015,Vijay_haah_fu_2016}, gauge theory~\cite{Pretko_U1_2017,Pretko_EM_2017}, quantum computing~\cite{Haah_code}, glasses and soft matter~\cite{Prem_Pretko_emergent,Prem_glassy_translationinvariant}, These exotic properties result from unusual mobility constraints whereby a fracton is a charge-like excitation which has restricted  mobility when in isolation, but which, nonetheless, can easily move in a subdimension of space when bound to an oppositely charged fracton in a dipolar bound state~\footnote{These are often referred to as type-I fractons}.

\begin{figure}[t]
\centering
\includegraphics[width=0.98\columnwidth]{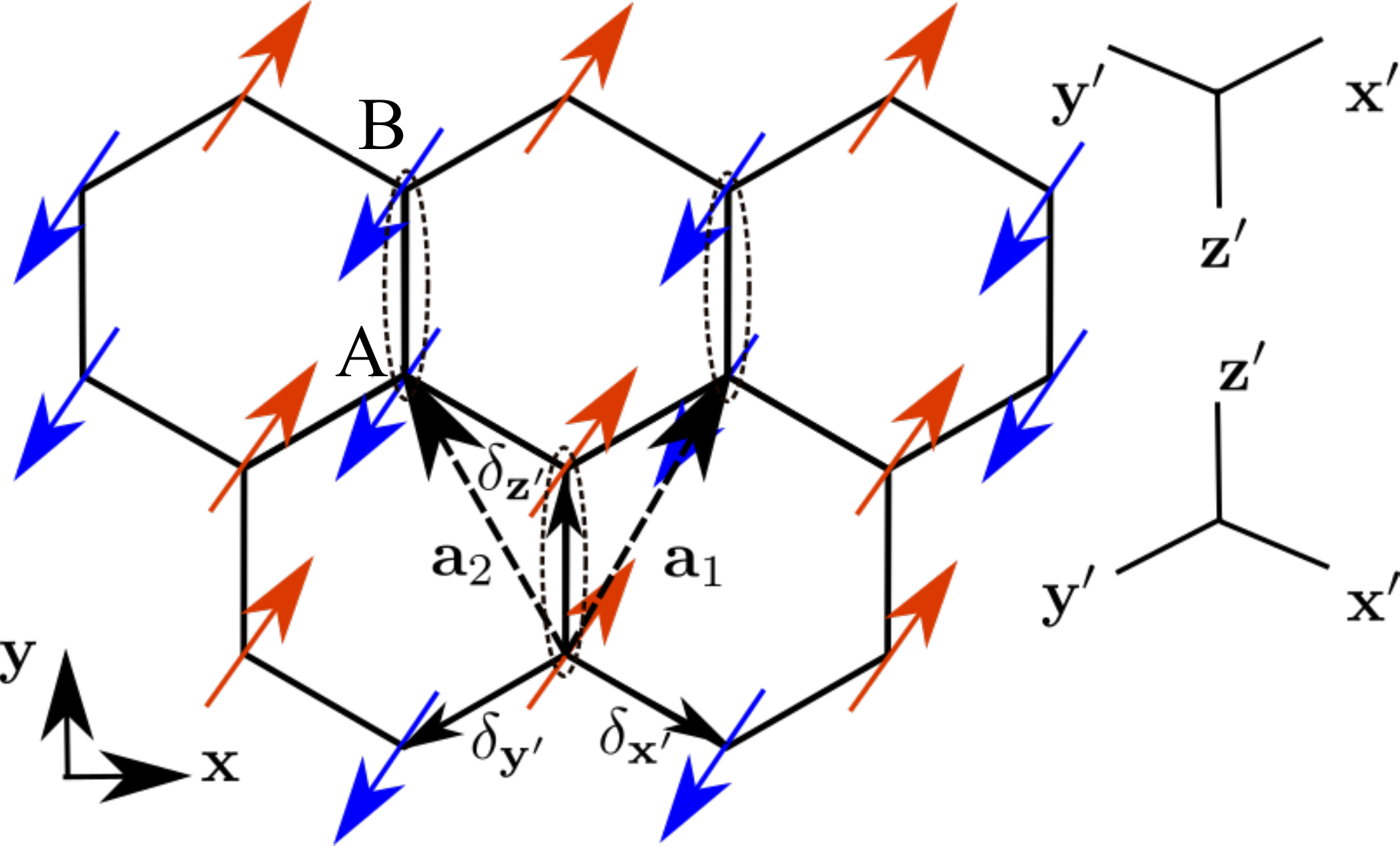}
\vspace*{-2mm}
\caption{Ground state of the Hamiltonian $\mathcal{H}_{\rm Ising}$ (Eq.~\eqref{Eq:spinmodel}) with one electron per site. The three bonds of the honeycomb lattice are denoted as $x^\prime,y^\prime$ and $z^\prime$ respectively ($x$ and $y$ are the Cartesian directions). The Hamiltonian describes an Ising magnet on a honeycomb lattice with antiferromagnetic exchange along the $x^\prime$ and $y^\prime$ bonds and ferromagnetic exchange along the $z^\prime$ bonds. The vectors $\delta_{x^\prime}$, $\delta_{y^\prime}$ and $\delta_{z^\prime}$ connect near-neighbor sites in the $x^\prime$, $y^\prime$ and $z^\prime$ directions, respectively. The unit cell of the honeycomb lattice with A and B sublattices is shown in the dotted region, and $a_1$ and $a_2$ are the primitive lattice vectors.}
\label{fig:mode1}
\end{figure}

\begin{figure*}[t]
\includegraphics[width=0.72\columnwidth]{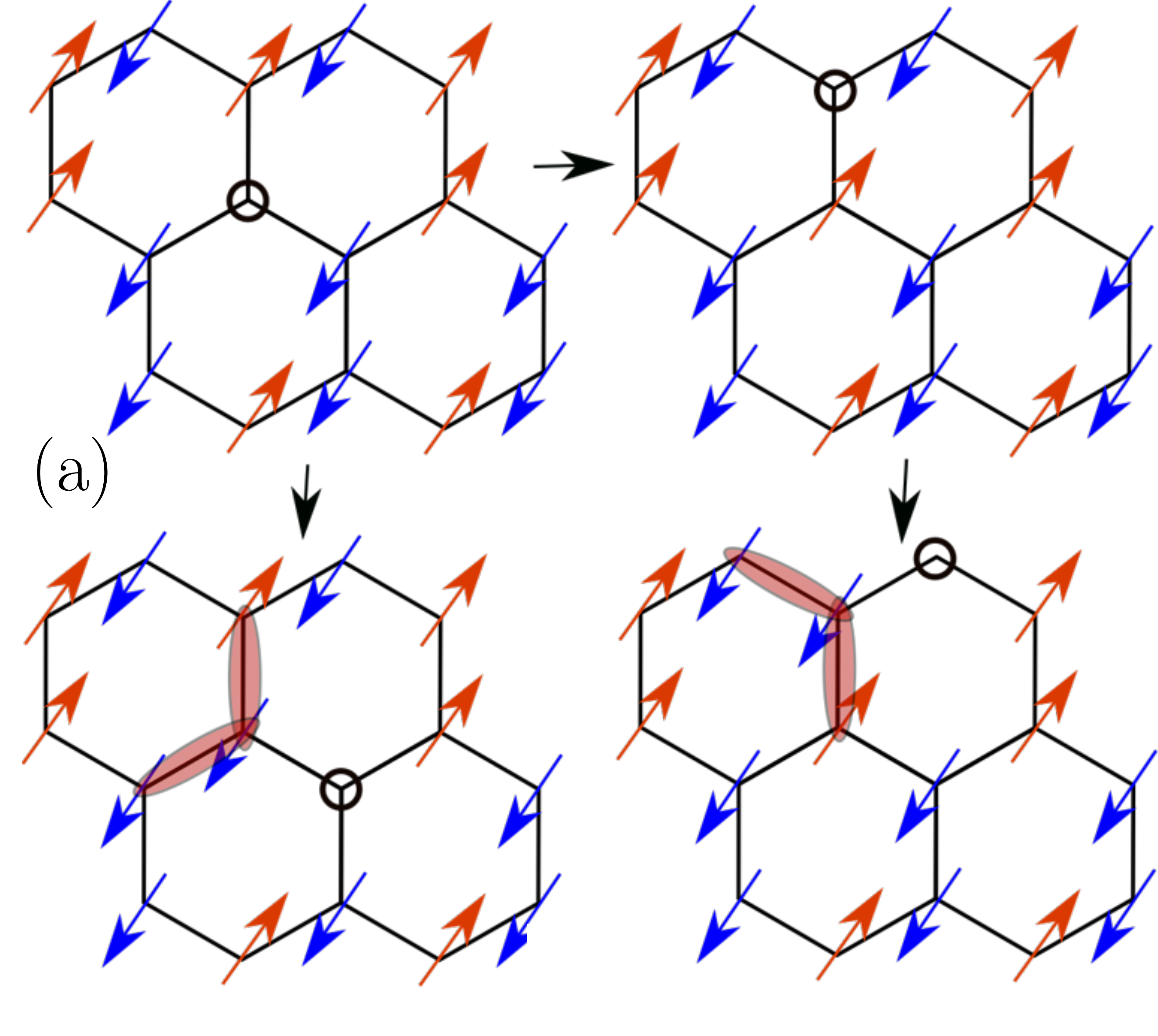} \hspace{4mm} \includegraphics[width=1.28\columnwidth]{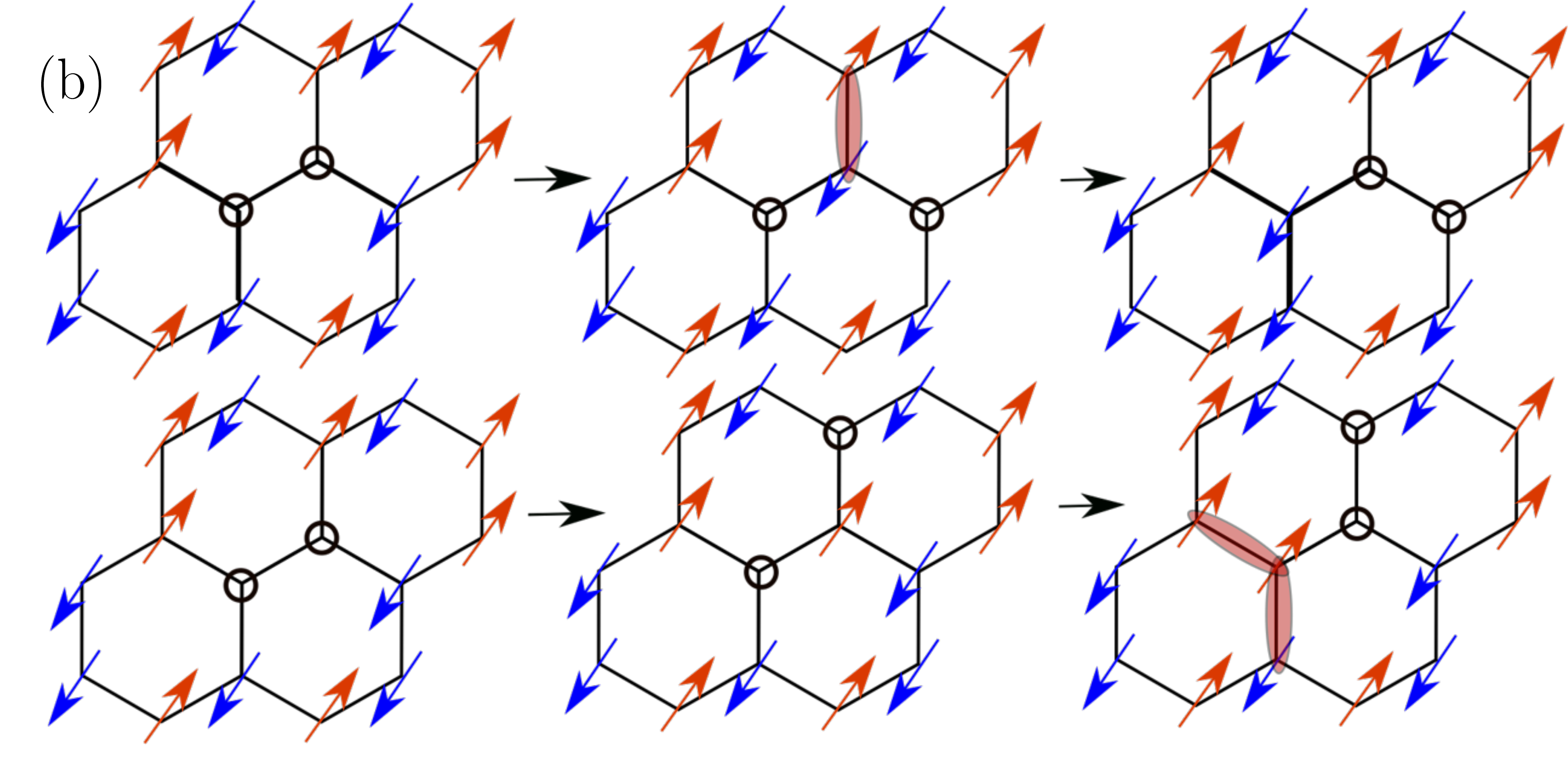}
\caption{\label{fig:singlehole} (a) An isolated hole can  move by one site only in the ferromagnetic direction but cannot move in the antiferromagnetic direction without frustrating the antiferromagnetic bonds. (b) A pair of holes on neighboring sites can move only along the antiferromagnetic $x-y$ direction. Note that a pair of holes on neighboring sites connected by a bond in the $z^\prime$ direction cannot move without frustrating the background and as such will be localized.}
\label{fig:holes}
\end{figure*}

The realization of these aberrant mobility constraints and subsystem symmetry in physical systems is, however, not naturally available, stimulating many interesting theoretical proposals~\cite{lineon_planon_hermele,foliated_fracton,screwdislocation_xcode}.  One particularly appealing approach is based on the idea that defect motion-induced frustration in an ordered background leads to emergent immobility constraints on the motion of defects themselves which in turn gives rise to fractonic quasiparticles, with hole-doped Ising antiferromagnets~\cite{TrugmanLoops,KaneLeeRead,SachdevHoleAFM,KivelsonHoleAFM,Intermediate_coupling_Barentzen,Sasha1,Sasha2,Grusdt,WohlfeldIsing1,WohlfeldIsing2}  being a prime example of this mechanism~\cite{Sous2020Fracton1,Sous2020Fracton2}.  Unlike other constructions proposed to realize fracton topological order (e.g.~\cite{halasz_CM}), this approach does not require an extensive number of locally conserved quantities (which results in an extensively degenerate ground state), making the physical realization of fracton conservation laws (without topological order) in quantum magnets significantly more accessible. However, the emergence of fracton-like quasiparticles in fully two-dimensional (2D) hole-doped Ising antiferromagnets has been demonstrated only in the asymptotic limit of $t \ll J$ (where $t$ is the hole hopping and $J$ is the Ising coupling)~\cite{Sous2020Fracton1}, where already at the leading order in $t^2/J$, the direction of the dipole moment is not conserved even though its magnitude is because hole pairs can rotate as they move in the 2D plane.

In this letter we overcome this limitation and propose an essentially exact physical realization of fractonic quasiparticles with subsystem dipolar symmetry in a 2D non-degenerate, ordered spin system with local two-spin interactions. The key ingredient in our proposal is the collinear antiferromagnetic order in which defect motion-induced frustration completely prevents hole pairs or dipoles from moving in the perpendicular direction, resulting in exact conservation of both the magnitude and direction of dipole moment in the asymptotic limit $t\ll J$. This further endows the system with a subsystem symmetry since dipole  motion is restricted to a one-dimensional (1D) submanifold of the system.  This symmetry manifests only along the antiferromagnetic direction, thus we denote it as `unidirectional' subsystem symmetry. Importantly, we find via exact diagonalization (ED) that this symmetry continues to hold quantitatively away from the perturbative limit when $t\gtrsim J$. We further suggest that interaction between hole pairs is weak implying continuity of these immobility constraints to small, finite concentrations.

\emph{Model}. Using a coupled spin chain construction to ferromagnetically couple antiferromagnetic Ising chains we construct a 2D ordered Ising magnet, in which we study doped holes. Specifically, we consider a model of holes doped into an Ising collinear antiferromagnet on the honeycomb lattice given by
\begin{equation}
    \mathcal{H} = -t\sum_{\textbf{r}_i, \delta_j, \sigma} \left( c_{\textbf{r}_i + \delta_j,\sigma}^\dagger c_{\textbf{r}_i, \sigma} + \text{h.c.}\right) + \mathcal{H}_\mathrm{Ising},
    \label{eq:hamiltonian}
\end{equation}
where $t$ denotes the hopping amplitude, $c^\dagger$ ($c$) are fermionic creation (annihilation) operators, and $\sigma \in \{\uparrow,\downarrow\}$ is the fermion spin.  The coordinates $\textbf{r}_i$ are defined by the sites of a Bravais lattice given by $\mathbf{r}_j=m_j \mathbf{a}_1+n_j\textbf{a}_2$ where $\mathbf{a}_1$, $\mathbf{a}_2$ denote the primitive lattice vectors of the honeycomb lattice. The basis has two sites referred as A and B. The vectors $\delta_{x^\prime},\delta_{y^\prime},\delta_{z^\prime}$ connect spins on nearest-neighbor sites in the three directions of the honeycomb lattice, see Fig.~\ref{fig:mode1}. A no double occupancy constraint, $n_{\textbf{r}_i} =\sum_\sigma c_{\textbf{r}_i, \sigma}^\dagger c_{\textbf{r}_i, \sigma} \leq 1$, $\forall \textbf{r}_i \in A,B$ is enforced on the Hilbert space. The Ising spin Hamiltonian $\mathcal{H}_\mathrm{Ising}$ is constructed by ferromagnetically coupling alternating sites on antiferromagnetic chains (this can be viewed as a model of a striped antiferromagnet in a brick-wall lattice). The ferromagnetic spin couplings are taken to be along the $\delta_{z^\prime}$ direction and antiferromagnetic spin couplings  along the $\delta_{x^\prime}$ and $\delta_{y^\prime}$ directions:
\begin{equation}
\mathcal{H}_\mathrm{Ising}= J\sum_{\textbf{r}_i} \left( S_{\textbf{r}_i+\delta_{x^\prime}}^z  S_{\textbf{r}_i}^z +S_{\textbf{r}_i+\delta_{y^\prime}}^z  S_{\textbf{r}_i}^z-S_{\textbf{r}_i+\delta_{z^\prime}}^z  S_{\textbf{r}_i}^z\right),
\label{Eq:spinmodel}
\end{equation}
where $S_{\textbf{r}_i}^z$ denotes the spin-$z$ operator. One ground state of $\mathcal{H}_\mathrm{Ising}$ is given by $| \psi_\mathrm{GS}\rangle =\prod_{ \textbf{r}_i.(\textbf{a}_1+\textbf{a}_2)\in 2\mathbb{Z}} c_{\textbf{r}_i,\downarrow}^\dagger  c_{\textbf{r}_i+\delta_{x^\prime},\downarrow}^\dagger  c_{\textbf{r}_i+\textbf{a}_1,\uparrow}^\dagger  c_{\textbf{r}_i+\textbf{a}_1+\delta_{z^\prime},\uparrow}^\dagger|0\rangle.$

A hole can be created (annihilated) on a given site by annihilating (creating) an electron on that site. For any given hole density, the total magnetization is conserved. Thus, we associate the removal (addition) of a fermion with spin $\sigma$ with the creation (annihilation) of a hole with spin $-\sigma$, as either amounts to a total net change of the magnetization of the entire system by $-\sigma$. Therefore the hole creation operator is given by $f_{\textbf{r}_i,\sigma}^\dagger=c_{\textbf{r}_i,-\sigma}$.  A hole can move to a neighboring site along the antiferromagnetic direction if the electron with antialigned spin on that site moves to the hole's original site. One can view this as a spin flip operation at the original hole site accompanied by hopping of the hole to the concerned neighbor site. The original hole site with a flipped spin is now in a ``wrong'' orientation with respect to its two remaining neighbors and thus we view this as defect creation. A hole dressed by such bosonic (spin wave) defects forms a magnetic polaron. We can represent a misaligned spin as a bosonic magnon defect for the sites $\textbf{r}_i.(\textbf{a}_1+\textbf{a}_2) \in 2\mathbb{Z}$ as  $b_{\textbf{r}_i}^\dagger =\sigma_{\textbf{r}_i}^-$, $b_{\textbf{r}_i+\delta_{x^\prime}}^\dagger = \sigma_{\textbf{r}_i+\delta_{x^\prime}}^+$, $b_{\textbf{r}_i+\delta_{y^\prime}}^\dagger=\sigma_{\textbf{r}_i+\delta_{y^\prime}}^+$, and $b_{\textbf{r}_i+\delta_{z^\prime}}^\dagger=\sigma_{\textbf{r}_i+\delta_{z^\prime}}^-$,
and for the sites  $\textbf{r}_i.(\textbf{a}_1+\textbf{a}_2) \in 2\mathbb{Z}+1$  as  $b_{\textbf{r}_i} =\sigma_{\textbf{r}_i}^+$, $b_{\textbf{r}_i+\delta_{x^\prime}}=\sigma_{\textbf{r}_i+\delta_{x^\prime}}^-$, 
$b_{\textbf{r}_i+\delta_{y^\prime}}=\sigma_{\textbf{r}_i+\delta_{y^\prime}}^-$, and $b_{\textbf{r}_i+\delta_{z^\prime}}=\sigma_{\textbf{r}_i+\delta_{z^\prime}}^+$ , where $\sigma_{\textbf{r}_i}^{\pm}$ are the Pauli Ladder operators. In contrast to motion along the antiferromagnetic direction, a single hole can move -- only by only one site -- in the ferromagnetic $z^\prime$ direction since there no wrongly aligned spin.

A unique characteristic of this model on the honeycomb lattice is that each site has two antiferromagnetic neighbors in the $x^\prime$ and $y^\prime$ directions and one ferromagnetic neighbor in the $z^\prime$ direction. This means that in order for a hole to move from a site to another in the $x^\prime$ or $y^\prime$ direction it must create a defect on the site of its departure or annihilate a defect on the site of its arrival. In contrast, hole hopping in the $z^\prime$ direction will not involve any defect creation. Thus, in the language of the hole and magnon defect operators our model in Eq.~\eqref{eq:hamiltonian} can be recast as 
%\begin{widetext}
%\begin{eqnarray}
%\mathcal{H}=&-& \sum_{\textbf{r}_i,\sigma} [t~(f_{\textbf{r}_i+\delta_{x^\prime},\sigma}^{\dagger}  f_{\textbf{r}_i,\sigma} (b_{\textbf{r}_i}^\dagger+ b_{\textbf{r}_i+\delta_{x^\prime}})+f_{\textbf{r}_i+\delta_{y^\prime},\sigma}^{\dagger}  f_{\textbf{r}_i,\sigma} (b_{\textbf{r}_i}^\dagger+ b_{\textbf{r}_i+\delta_{y^\prime}}))-t~f_{\textbf{r}_i+\delta_{z^\prime},\sigma}^{\dagger}  f_{\textbf{r}_i,\sigma}\\ \nonumber
%&-& g~(f_{\textbf{r}_i,\sigma}^{\dagger}  f_{\textbf{r}_i+\delta_{x^\prime},\sigma} (b_{\textbf{r}_i+\delta_{x^\prime}}^\dagger+ b_{\textbf{r}_i})+f_{\textbf{r}_i,\sigma}^{\dagger}f_{\textbf{r}_i+\delta_{y^\prime},\sigma}   (b_{\textbf{r}_i+\delta_{y^\prime}}^\dagger+ b_{\textbf{r}_i}))-t~f_{\textbf{r}_i,\sigma}^{\dagger}  f_{\textbf{r}_i+\delta_{z^\prime},\sigma} ] + H_{Ising}.
%\end{eqnarray}
\begin{eqnarray}
\mathcal{H}=&-&t \sum_{\textbf{r}_i} \big(f_{\textbf{r}_i+\delta_{z^\prime},\sigma}^{\dagger}  f_{\textbf{r}_i,\sigma}+  \mathrm{h.c.}  \big) \nonumber  \\  
&-&t \sum_{\textbf{r}_i} \Bigg[ f_{\textbf{r}_i+\delta_{x^\prime},\sigma}^{\dagger}  f_{\textbf{r}_i,\sigma} (b_{\textbf{r}_i}^\dagger+ b_{\textbf{r}_i+\delta_{x^\prime}}) +  \mathrm{h.c.} \Bigg] \nonumber \\  
&-&t \sum_{\textbf{r}_i} \Bigg[ f_{\textbf{r}_i+\delta_{y^\prime},\sigma}^{\dagger}  f_{\textbf{r}_i,\sigma} (b_{\textbf{r}_i}^\dagger+ b_{\textbf{r}_i+\delta_{y^\prime}}) +  \mathrm{h.c.} \Bigg] \nonumber  \\ 
&+& \mathcal{H}_\mathrm{Ising},
\label{t-Jz_model}
\end{eqnarray}%{\substack{\textbf{r}_i \in A \\ \hat{y}.\textbf{r}_i \in even}}
%\end{widetext}
%where $\omega_b$ is the energy cost to create a boson, so $\omega_b=nJ$ when a bosonic defect is created with $n-$neighboring ising spins. 
In Eq.~\eqref{t-Jz_model} a no-double occupancy constraint at each lattice position is imposed, as there can be either a defect or a spin in a given lattice site. We ascribe an effective charge degree of freedom to the dressed hole in terms of its spin density: $q_{\textbf{r}_i}=  f^\dagger_{\textbf{r}_i,\alpha} \tau^z_{\alpha,\alpha} f_{\textbf{r}_i,\alpha}$, where $\tau_z$ is the Pauli-$z$ matrix in the hole spin flavor space. The total charge $Q=\sum_iq_{\textbf{r}_i}$ is a global conserved quantity, $[\mathcal{H},Q]=0$.

\begin{figure}[t]
\centering
\includegraphics[width=\columnwidth]{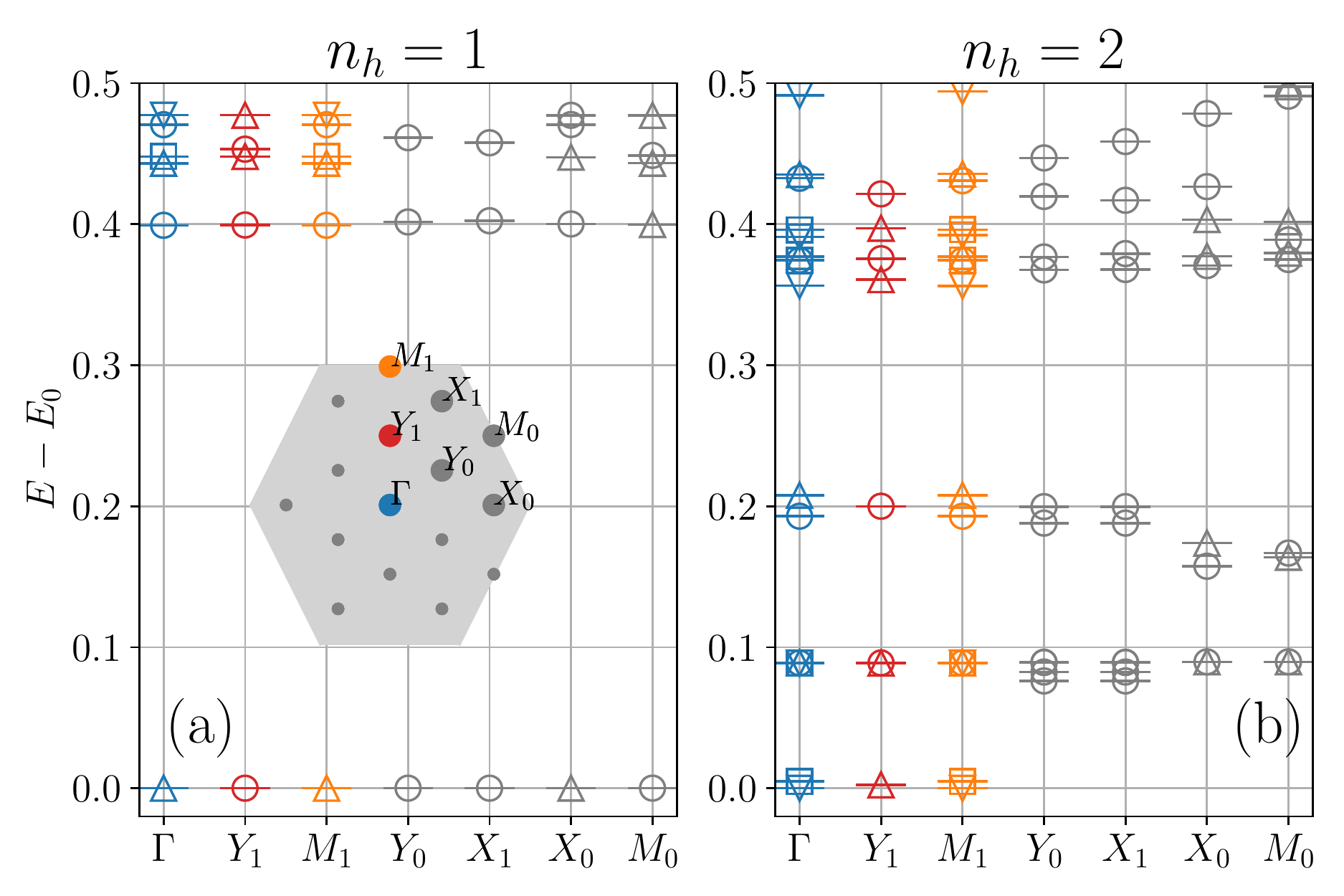}
\includegraphics[width=\columnwidth]{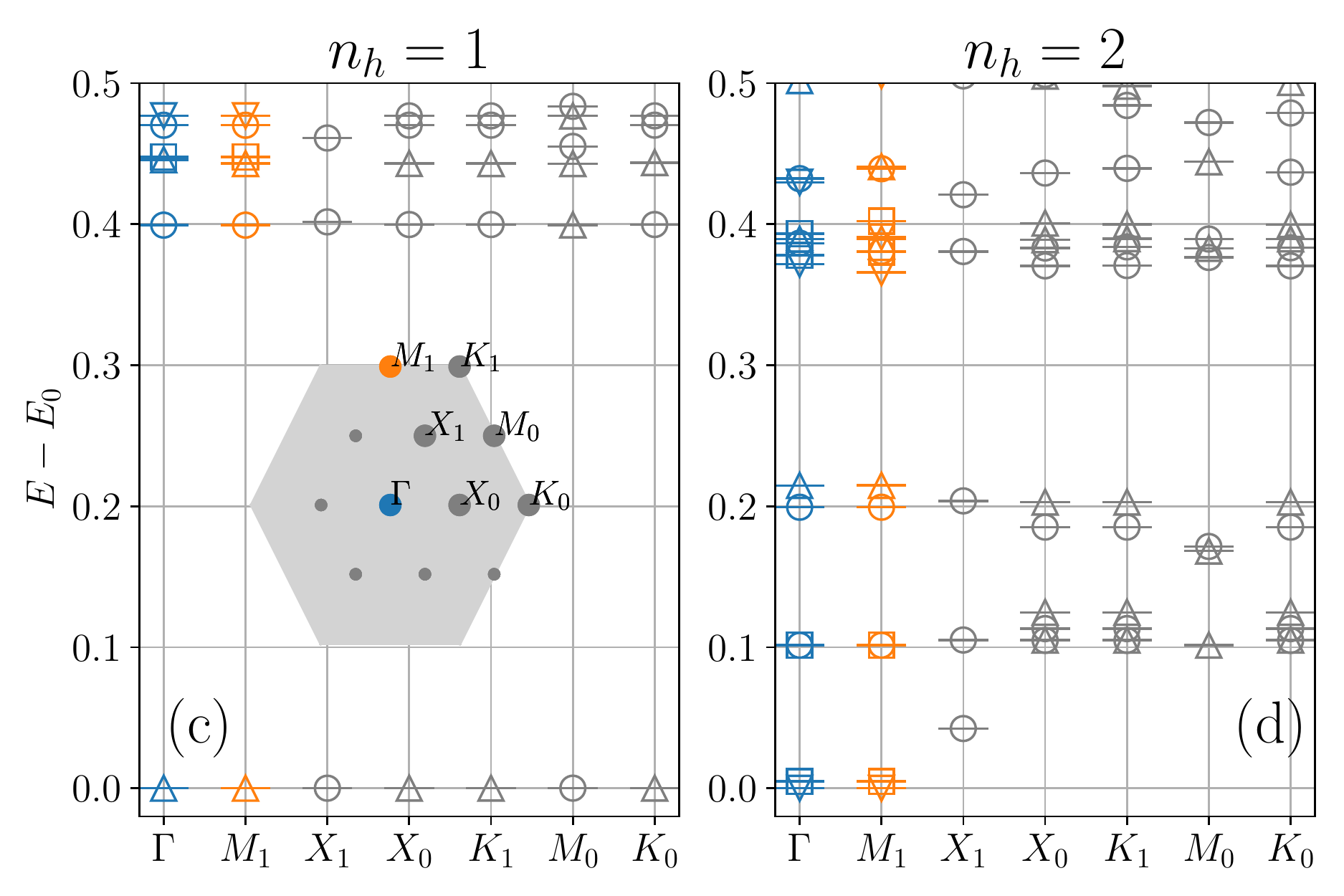}
\caption{Energy spectrum as a function of momentum from  ED on an $N=32$ ((a) and (b)) and $N=24$ ((c) and (d)) cluster for $t=1.0$ and $J=0.4$. Results for a single hole are shown in (a) and (c), and results for two holes are shown in (b) and (d). The insets depict the momenta resolved by the simulation cluster. The symbols of the energy levels denote different point group representations. We observe a flat spectrum in both momentum directions for $n_h=1$ and a flat spectrum in the $k_y$ direction only for $n_h=2$. This implies full localization of the single hole and localization of the two holes along the $\delta_{z^\prime}$ direction  only.}
\label{fig:edspectra}
\end{figure}

\emph{Single hole}. We first consider the single-hole sector $n_h \equiv\sum_{\textbf{r}_i,\sigma}
f_{\textbf{r}_i,\sigma}^\dagger f_{\textbf{r}_i,\sigma}=1$. We find an effective Hamiltonian for the single hole in the limit $t/J \ll 1$ by treating the hole-boson coupling term in Eq.~\eqref{t-Jz_model} perturbatively. The effective Hamiltonian, up to second order, is given by
\begin{equation}
    h^{2nd}_{1h}=-\frac{2t^2}{J}\sum_{\textbf{r}_i,\sigma}
f_{\textbf{r}_i,\sigma}^\dagger f_{\textbf{r}_i,\sigma}-t \sum_{\textbf{r}_i,\sigma}
\big( f_{\textbf{r}_i+\delta_{z^\prime},\sigma}^\dagger f_{\textbf{r}_i,\sigma} + \mathrm{h.c.} \big).
\label{single_particle_eqn}
\end{equation}
The first term in this expression reflects the localization of single holes where $2t^2/J$ is the formation energy associated with creating an immobile polaron. This can be understood as follows. Since hole motion in the $x^\prime$ and $y^\prime$ directions frustrates the antiferromagnetic bonds, the hole must retrace its path and return to its original site in order to heal the background~\cite{Sous2020Fracton1}, thus a single hole cannot move along the antiferromagnetic direction. This is represented pictorially in Fig.~\ref{fig:singlehole} where we demonstrate that a single hole can hop only between two sites in the $z^\prime$ direction via the second term in Eq.~\eqref{single_particle_eqn}, but cannot hop in the $x^\prime$ and $y^\prime$ directions without creating misaligned spins even following a move in $x^\prime$ direction, and thus the hole is localized on the $z^\prime$ bond connecting the two sites.  A single hole can only move away from its original site via virtual processes involving closed loops known as Trugman loops~\cite{TrugmanLoops}. In contrast to the case of a square-lattice antiferromagnet where Trugman loops first appear at sixth order in perturbation theory, Trugman loops in our model first appear only at fifteenth order.  Thus, we expect fracton physics to persist in a paramatrically larger regime in $t/J$. Furthermore, away from the perturbative limit, we expect these closed loops to be energetically much more costly and thus their contribution to hole motion to be negligible~\cite{WohlfeldIsing2}. To confirm this behavior beyond the limits of applicability of perturbation theory, we use ED~\cite{Wietek2018} for the model with $J/t=0.4$ on $N=24$ and $N=32$ site clusters. Figure~\ref{fig:edspectra}(a), (c) show the energy spectrum as a function of momentum in the Brillouin zone for $n_h=1$. We observe nearly exact degeneracy of the spectrum at all momenta. This fully flat degeneracy in momentum space indicates localization in real space, as expected from our arguments for the spin polaron~\footnote{The spread of energy $\Delta = \max( |E_1(\Gamma) - E_0(\Gamma)|, |E_0(M_1) - E_0(\Gamma) |, | E_1(M_1) - E_0(\Gamma)|) $ in this (quasi-)degenerate multiplet is remarkably small: $\Delta/t \approx 4.4 \times 10^{-5}$ for $N=24$ and $\Delta/t \approx 6.6 \times 10^{-5}$ for $N=32$. The small increase of $\Delta$ from $N=24$ to $N=32$ is attributed to non-trivial cluster geometry effects, which arise if the spanning vectors of the embedded simulation clusters are not proportional to one another. The gap to excitations is found to be exactly of the order of the Ising interaction $J/t=0.4$.}.

\emph{Two holes}. Next, we consider the two-hole sector $n_h = \sum_{\textbf{r}_i,\sigma} f_{\textbf{r}_i,\sigma}^\dagger f_{\textbf{r}_i,\sigma}=2$. We derive an  effective Hamiltonian for two holes perturbatively, which, up to second order in $t/J$ is given by:
\begin{widetext}
\begin{eqnarray}
h^{2nd}_{2h}&=&   -\frac{2t^2}{J}\sum_{\textbf{r}_i,\sigma}
f_{\textbf{r}_i,\sigma}^\dagger f_{\textbf{r}_i,\sigma}
-t \sum_{\textbf{r}_i,\sigma} \big(
f_{\textbf{r}_i+\delta_{z^\prime},\sigma}^\dagger f_{\textbf{r}_i,\sigma}+ \mathrm{h.c.}\big)+\frac{4t^2}{J} \sum_{\substack{\textbf{r}_i,\sigma,k\in\{x^\prime,y^\prime\}}}n_{\textbf{r}_i,\sigma} n_{\textbf{r}_i+\delta_k,-\sigma} +\frac{8t^2}{J}\sum_{\substack{\textbf{r}_i,\sigma}}n_{\textbf{r}_i,\sigma}n_{\textbf{r}_i+\delta_{z^\prime},\sigma} \nonumber \\
&&-\frac{2t^2}{J} \sum_{\substack{\textbf{r}_i,\sigma}} \Big( f_{\textbf{r}_i+\delta_{x^\prime},\sigma}^\dagger f_{\textbf{r}_i,-\sigma}^\dagger f_{\textbf{r}_i,\sigma} f_{\textbf{r}_i+\delta_{y^\prime},-\sigma}  + f_{\textbf{r}_i-\textbf{a}_1+\textbf{a}_2,-\sigma}^\dagger f_{\textbf{r}_i+\delta_{x^\prime},\sigma}^\dagger f_{\textbf{r}_i,\sigma} f_{\textbf{r}_i+\delta_{y^\prime},-\sigma}  +\mathrm{h.c.} \Big) \nonumber \\
&&- \frac{2t^2}{J}\sum_{\substack{\textbf{r}_i,\sigma}} \Big( f_{\textbf{r}_i+\delta_{y^\prime},\sigma}^\dagger f_{\textbf{r}_i,-\sigma}^\dagger f_{\textbf{r}_i,\sigma} f_{\textbf{r}_i+\delta_{x^\prime},-\sigma}  + f_{\textbf{r}_i+\textbf{a}_1-\textbf{a}_2,\sigma}^\dagger f_{\textbf{r}_i+\delta_{y^\prime},-\sigma}^\dagger f_{\textbf{r}_i,\sigma} f_{\textbf{r}_i+\delta_{x^\prime},-\sigma}  + \mathrm{h.c.} \Big).
\label{two_particle_eqn}
\end{eqnarray}
\end{widetext}
Here, the first and second terms correspond to single-hole processes derived in Eq.~\eqref{single_particle_eqn}, while the third and fourth terms correspond to near-neighbor repulsive density-density interactions between  magnetic polarons, and the fifth and sixth terms correspond to pair hopping~\cite{SousScRep,SousBipolaron} along the $x^\prime-y^\prime$ direction. This effective Hamiltonian shows that two neighboring holes oriented along the $x^\prime-y^\prime$ line can move in a bound state via a second-order process in which  one hole hops by creating a bosonic defect which is then absorbed by the second hole {\em only} if it follows its partner  along the $x^\prime$ or $y^\prime$ directions. In contrast, hole pairs cannot move along the $z^\prime$-direction because any such motion frustrates the antiferromagnetic bonds, in which case the holes have no option but to retrace their paths to their original non-frustrating configuration. One can visualize these processes pictorially in  Fig.~\ref{fig:holes} which shows that two holes can move only together and only in the antiferromagnetic $x^\prime$-$y^\prime$ direction, but not through the ferromagnetic $z^\prime$ direction~\footnote{Any two non-neighboring holes behave like two isolated single holes.  Here a hole can hop with  amplitude  $t$ in the $z^\prime$ direction, but no further hopping is possible without creating bosonic defects, which, barring closed loops, can be removed only via self-retracing motion of the hole.}.  This picture is confirmed via ED for the model with $J/t=0.4$ on $N=24$ and $N=32$ site clusters. Figure~\ref{fig:edspectra}(b), (d) shows the energy spectrum as a function of momentum in the Brillouin zone for  $n_h=2$. We observe that the lowest-energy states in the different momentum sectors are nearly exactly degenerate {\em only} along the $k_y$-direction of the Brillouin zone, indicating localization of states solely along the  $\delta_{z^\prime}$ direction in real space~\footnote{Again the difference between the energy levels is remarkably small. For $N=24$ the energy difference between the (quasi-)degenerate states at $\Gamma$ and $M_1$ is $\Delta /t \approx 4.83\cdot 10^{-3}$ while for $N=32$ the energy difference between states at $\Gamma$, $M_1$ and $Y_1$ is $\Delta /t \approx 4.79\cdot 10^{-3}$.  The energy differences between the lowest energy states $E_0(\mathbf{k})$ at $\Gamma$ and $M_1$ are vanishingly small; $|E_0(\Gamma) - E_0(M_1) | \approx 4.72 \cdot 10^{-7}$ for $N=24$ and $|E_0(\Gamma) - E_0(M_1) | \approx 4.28 \cdot 10^{-12}$ for $N=32$.}.

\begin{figure}[b]
\centering
\includegraphics[width=\columnwidth]{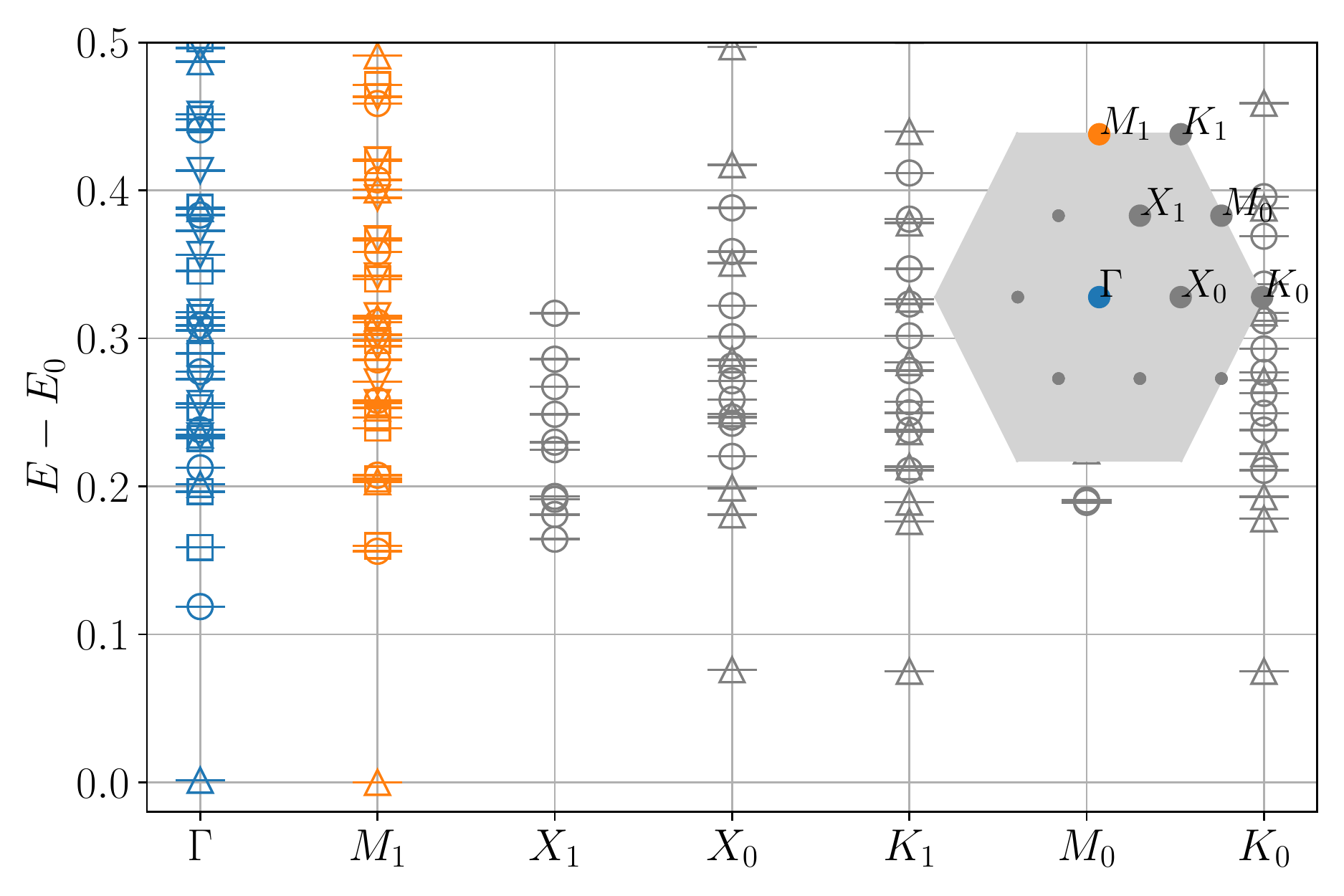}
\caption{Energy spectrum as a function of momentum for $n_h=4$ holes from ED on an $N=24$ cluster for $t=1.0$ and $J=0.4$. The momenta $\Gamma$ and $M_1$ are degenerate up to a difference of $\Delta/t = 1.4\cdot 10^{-3}$, indicating localization along the $k_y$ direction.}
\label{fig:edspectra_4}
\end{figure}

The phenomenology of one- and two-hole states in our model implies an emergent unidirectional subsystem symmetry   along the antiferromagnetic direction with conservation of both charge and dipole moment (defined as $D=\sum_i q_{\textbf{r}_i} \textbf{r}_i$). Here  a single hole forms a spin polaron which is almost perfectly localized up to very high order in perturbation theory mimicking a fracton, while two near-neighbor holes form a spin bipolaron which moves with ease along an antiferromagnetic 1D submanifold of the system while conserving dipole moment exactly like a dipole.   The effective Hamiltonian in Eq.~\eqref{two_particle_eqn} manifestly conserves the total charge. The conservation of dipole moment, $[D,h_{2h}^{2nd}]=0$~\footnote{This commutation relation holds in a system with open boundary conditions.}, can be seen from Eq.~\eqref{two_particle_eqn} which shows that hole pairs can only move together whilst conserving their relative separation and cannot rotate~\footnote{Note that a single hole can move along the $z^\prime$ direction according to the second term of Eq.~\eqref{two_particle_eqn}, but the dipole moment is still conserved as it does not change the Bravais lattice coordinate.}. We note that unlike the case of the square-lattice antiferromagnet~\cite{Sous2020Fracton1}, there is no need to impose external energetic constraints to realize fracton physics in our model, which appears to hold robustly even away from the perturbative limit $t/J \ll 1$ as seen in ED.  Furthermore, our results will continue to hold even if the magnitude of $J$ along the antiferromagnetic direction is different from that along the ferromagnetic direction, making physical realization more readily accessible.

\emph{Finite density of holes}. Having established unidirectional dipole symmetry, we address the question of dipole-dipole interactions at finite hole concentrations. We study numerically via ED $n_h = 4$ holes in a $N=24$-site cluster (large clusters are beyond reach). Figure~\ref{fig:edspectra_4}  shows a nearly degenerate spectrum along the $k_y$ momentum direction corresponding to the ferromagnetic direction $\delta_{z^\prime}$ in real space suggesting that dipole conservation may persist at finite hole densities.  To investigate interactions between dipolar pairs we compute the pair-pair binding energy $E_{\text{pair}}$ defined as
\begin{equation}
    E_{\text{pair}} = E_0(n_h=4) - 2 E_0 (n_h=2) + E_0 (n_h=0).
\end{equation}
For the $N=24$ site cluster, at $J/t = 0.4$ we find $E_{\text{pair}}/t=0.13$, suggesting that dipole-dipole interaction is repulsive. However, it is not clear to what extent these results are sensitive to finite-size effects.  We argue that current indications suggest that dipole symmetry may play an important role  at finite hole concentrations, and hope to address this in future work.

\emph{Conclusion}. We considered the physics of fractons and dipoles emergent in a hole-doped collinear antiferromagnets on a honeycomb lattice. By means of analytical arguments and ED, we showed that individual holes are completely localized in the 2D system, while near-neighbor hole pairs form dipolar lineons which can move freely only along the antiferromagnetic direction.  These observations reflect an emergent quasi-exact unidirectional subsystem symmetry along the antiferromagnetic direction. These results were obtained for an Ising magnet, but, based on perturbative arguments~\cite{Sous2020Fracton1}, we expect a sufficiently small spin exchange $J_\perp$ to not affect our results significantly~\cite{Sasha3,Sasha4}.   Our results indicate that dipole symmetry is a robust feature in the limit of a single and a pair of doped holes and may persist to finite hole concentrations where dipole-dipole interactions are non-trivial and have implications that will be the subject of future work.  Another promising future direction involves using a coupled plane construction analogous to our approach to engineer a three-dimensional model with exact subsystem symmetry along lines or planes. We hope that such approaches will enable the study of the exotic properties of fractons such as their unusual dynamical behavior in simple, potentially accessible spin systems.

\begin{acknowledgements}
We acknowledge useful discussions with M. Bukov, A.~L.~Chernyshev, J.~Romhanyi,  C.~Xu, and especially with M.~Berciu, V.~Calvera, K.-S. Kim, A.~Prem, K.~Wohlfeld and  H.~Yan. S.~S. acknowledges support from Science and Engineering Research Board (Department of Science and Technology) Govt. of India, under grant no. SRG/2020/001525 and an internal start up grant from Indian Institute of Science Education and Research, Tirupati. J.~S. acknowledges support from the Gordon and Betty Moore Foundation’s EPiQS Initiative through Grant GBMF8686 at Stanford University and the National Science Foundation (NSF) Materials Research Science and Engineering Centers (MRSEC) program through Columbia University (where this work was initiated) in the Center for Precision Assembly of Superstratic and Superatomic  Solids under Grant No. DMR-1420634.  J.~S. also acknowledges the hospitality of the Center for Computational Quantum Physics (CCQ) at the Flatiron Institute. The Flatiron Institute is a division of the Simons Foundation. Exact diagonalization calculations were performed using the Hydra software package~\cite{hydra}.
\end{acknowledgements}

\catcode`'=9
\catcode``=9
\bibliography{bib}

\end{document}